\documentclass[sigconf]{acmart} 

\AtBeginDocument{%
  \providecommand\BibTeX{{%
    \normalfont B\kern-0.5em{\scshape i\kern-0.25em b}\kern-0.8em\TeX}}}

\copyrightyear{2024}
\acmYear{2024}
\setcopyright{licensedusgovmixed}\acmConference[ICPE '24 Companion]{Companion of the 15th ACM/SPEC International Conference on Performance Engineering}{May 7--11, 2024}{London, United Kingdom}
\acmBooktitle{Companion of the 15th ACM/SPEC International Conference on Performance Engineering (ICPE '24 Companion), May 7--11, 2024, London, United Kingdom}
\acmDOI{10.1145/3629527.3651428}
\acmISBN{979-8-4007-0445-1/24/05}

\usepackage{xspace}
\usepackage{xcolor}
\usepackage{amsmath, amsfonts, amsthm} 
\usepackage{graphicx}
\usepackage{float} 
\usepackage{hyperref}
\usepackage{wrapfig}
\usepackage{mathrsfs}
\usepackage{multicol}
\usepackage{balance}
\usepackage{mathtools}
\usepackage{tikz}

\usepackage{tabularx} 
\usepackage{adjustbox} 
\usepackage{multirow} 

\usepackage{complexity}
\usepackage{relsize}
\usepackage{graphicx}
\usepackage{epsfig}
\usepackage{flushend}
\usepackage{epstopdf}
\usepackage{longtable}
\usepackage{stmaryrd}
\usepackage{pstricks}
\usepackage{pst-tree}
\usepackage{epstopdf}
\usepackage{lipsum}
\usepackage{color}
\usepackage{framed}
\usepackage[normalem]{ulem}
\usepackage{float}
\usepackage{caption}
\usepackage{fancyvrb}
\usepackage{bm}

\usepackage{algorithmic}
\usepackage{algorithm}

\usepackage{textcomp}
\usepackage{fancyhdr}

\usepackage{pifont} 

\usepackage{multicol}
\usepackage{wrapfig}
\usepackage{subfig}
\usepackage{balance}

\begin{document}

\title{Evaluating Emerging AI/ML  Accelerators: IPU, RDU, and NVIDIA/AMD GPUs }



\thanks{$^*$Work done during an internship at Pacific Northwest National Laboratory}
\author{Hongwu Peng$^*$} 
\affiliation{
\institution{University of Connecticut}
\city{Storrs}
\state{CT}
\country{USA}
}
\email{hongwu.peng@uconn.edu}

\author{Caiwen Ding}
\affiliation{
\institution{University of Connecticut}
\city{Storrs}
\state{CT}
\country{USA}
}
\email{caiwen.ding@uconn.edu}

\author{Tong Geng}
\affiliation{
\institution{University of Rochester}
\city{Rochester}
\state{NY}
\country{USA}
}
\email{tgeng@ur.rochester.edu}

\author{Sutanay Choudhury}
\affiliation{
\institution{Pacific Northwest National Laboratory}
\city{Richland}
\state{WA}
\country{USA}
}
\email{sutanay.choudhury@pnnl.gov}

\author{Kevin Barker}
\affiliation{
\institution{Pacific Northwest National Laboratory}
\city{Richland}
\state{WA}
\country{USA}
}
\email{kevin.barker@pnnl.gov}

\author{Ang Li}
\affiliation{
\institution{Pacific Northwest National Laboratory}
\city{Richland}
\state{WA}
\country{USA}
}
\email{ang.li@pnnl.gov}

\renewcommand{\shortauthors}{Hongwu Peng, Caiwen Ding, Tong Geng, Sutanay Choudhury, Kevin Barker, \& Ang Li}


\begin{abstract}


The relentless advancement of artificial intelligence (AI) and machine learning (ML) applications necessitates the development of specialized hardware accelerators capable of handling the increasing complexity and computational demands. Traditional computing architectures, based on the von Neumann model, are being outstripped by the requirements of contemporary AI/ML algorithms, leading to a surge in the creation of accelerators like the Graphcore Intelligence Processing Unit (IPU), Sambanova Reconfigurable Dataflow Unit (RDU), and enhanced GPU platforms. These hardware accelerators are characterized by their innovative data-flow architectures and other design optimizations that promise to deliver superior performance and energy efficiency for AI/ML tasks.

This research provides a preliminary evaluation and comparison of these commercial AI/ML accelerators, delving into their hardware and software design features to discern their strengths and unique capabilities. By conducting a series of benchmark evaluations on common DNN operators and other AI/ML workloads, we aim to illuminate the advantages of data-flow architectures over conventional processor designs and offer insights into the performance trade-offs of each platform. The findings from our study will serve as a valuable reference for the design and performance expectations of research prototypes, thereby facilitating the development of next-generation hardware accelerators tailored for the ever-evolving landscape of AI/ML applications. Through this analysis, we aspire to contribute to the broader understanding of current accelerator technologies and to provide guidance for future innovations in the field.

\end{abstract}



\maketitle

\begin{table*}[t!]
\caption{Device information of Graphcore IPU, Sambanova RDU, and Nvidia/AMD GPU}
\centering
\resizebox{0.98\linewidth}{!}{
\label{tab:device_info}
\begin{tabular}{cccccccccccc}
\hline
Device        & \begin{tabular}[c]{@{}c@{}}TSMC\\ Process\end{tabular} & \begin{tabular}[c]{@{}c@{}}Die size\\ (mm\textasciicircum{}2)\end{tabular} & Transistors & Archi.                                                  & \begin{tabular}[c]{@{}c@{}}On-chip SRAM\\ (MB)\end{tabular}           & \begin{tabular}[c]{@{}c@{}}Off-chip\\ Memory\end{tabular}        & \begin{tabular}[c]{@{}c@{}}Clk\\ (GHz)\end{tabular} & \begin{tabular}[c]{@{}c@{}}FP64\\ (TFLOPs)\end{tabular}                      & \begin{tabular}[c]{@{}c@{}}FP32\\ (TFLOPs)\end{tabular}                      & \begin{tabular}[c]{@{}c@{}}FP16\\ (TFLOPs)\end{tabular} & Power \\ \hline \hline
IPU   GC2000  & 7nm                                                    & 823                                                                        & 59.4 B      & \begin{tabular}[c]{@{}c@{}}IPU\\ (MIMD)\end{tabular}    & 900@Scratchpad                                                        & \begin{tabular}[c]{@{}c@{}}448 GB@DRAM\\ 20 GB/s\end{tabular}    & 1.33                                                & \textbackslash{}                                                             & 62.5 (AMP)                                                                   & 250 (AMP)                                               & 165 W \\ \hline
SN10 RDU      & 7nm                                                    & N/A                                                                        & 40 B        & \begin{tabular}[c]{@{}c@{}}RDU\\ (CGRA)\end{tabular}    & 320@PMU                                                               & \begin{tabular}[c]{@{}c@{}}1.5 TB@DRAM\\ 100 GB/s\end{tabular}   & N/A                                                 & \textbackslash{}                                                             & \textbackslash{}                                                             & 325                                                     & N/A   \\ \hline
Nvidia   V100 & 12nm                                                   & 815                                                                        & 21.1 B      & \begin{tabular}[c]{@{}c@{}}Volta\\ (SIMT)\end{tabular}  & \begin{tabular}[c]{@{}c@{}}10.2@L1 Cache\\ 6.1@L2 Cache\end{tabular}  & \begin{tabular}[c]{@{}c@{}}16 GB@HBM2\\ 1.13 TB/s\end{tabular}   & 1.41                                                & 7.8 (CUDA Core)                                                              & 15.7 (CUDA Core)                                                             & 125 (Tensor Core)                                       & 250 W \\ \hline
Nvidia   A100 & 7nm                                                    & 826                                                                        & 54.2 B      & \begin{tabular}[c]{@{}c@{}}Ampere\\ (SIMT)\end{tabular} & \begin{tabular}[c]{@{}c@{}}24.6@L1 Cache\\ 40.9@L2 Cache\end{tabular} & \begin{tabular}[c]{@{}c@{}}40GB@HBM2\\ 1.6 TB/s\end{tabular}     & 1.6                                                 & \begin{tabular}[c]{@{}c@{}}9.7 (CUDA Core)\\ 19.5 (Tensor Core)\end{tabular} & 19.5 (CUDA Core)                                                             & 312 (Tensor Core)                                       & 250 W \\ \hline
AMD MI100     & 7nm                                                    & 750                                                                        & 25.6 B      & \begin{tabular}[c]{@{}c@{}}CDNA\\ (SIMT)\end{tabular}   & \begin{tabular}[c]{@{}c@{}}1.92@L1 Cache\\ 8@L2 Cache\end{tabular}    & \begin{tabular}[c]{@{}c@{}}32 GB@HBM2\\ 1.23 TB/s\end{tabular}   & 1.5                                                 & 11.5 (CU Core)                                                               & \begin{tabular}[c]{@{}c@{}}23.1 (CU Core)\\ 46.14 (Matrix core)\end{tabular} & 184.57 (Matrix Core)                                    & 290 W \\ \hline
\end{tabular}
}
\label{tab:hardware_info}
\end{table*}




\section{Introduction}

The rapid expansion of artificial intelligence (AI) and machine learning (ML) applications has led to a paradigm shift in computational hardware design. Traditional von-Neumann architectures, which have served as the backbone of computing for decades, are increasingly challenged by the demands of modern AI/ML workloads. These workloads often involve complex operations such as matrix multiplications, convolutions, and graph processing, which can be highly parallelizable but are bottlenecked by the data transfer constraints inherent in the von-Neumann architecture. To address these challenges, there has been a surge in the development of specialized hardware accelerators that aim to optimize the performance of AI/ML tasks through innovative architectural designs and execution models.

Among the emerging contenders in the field of AI/ML accelerators, Graphcore's Intelligence Processing Unit (IPU)~\cite{graphcore_GC200} and Sambanova's Reconfigurable Dataflow Unit (RDU)~\cite{prabhakar2018plasticine} stand out for their unique approach to hardware acceleration. These platforms leverage data-flow architectures, which are fundamentally different from the von-Neumann architecture, to enable more efficient computation for AI/ML workloads. By aligning the hardware design with the data-centric nature of AI/ML algorithms, these accelerators promise significant gains in performance and energy efficiency.

In addition to these specialized data-flow accelerators, Graphics Processing Units (GPUs) have also been at the forefront of AI/ML acceleration. With their highly parallel structure and robust ecosystem, GPUs continue to evolve with features such as Tensor Cores and enhanced memory hierarchies to better support the intensive computational demands of AI/ML applications.

In this study, we aim to provide a comprehensive evaluation and comparison of these commercial AI/ML accelerators. Our research delves into the architectural intricacies of the Graphcore IPU, Sambanova RDU, and various GPU platforms, examining their system design, memory hierarchy, computing resources, and programming models. By conducting a series of benchmark evaluations across a range of DNN operators, we seek to uncover the strengths and limitations of each platform, offering insights into their lower-precision floating-point numerical performance characteristics and suitability for different AI/ML tasks.

Our findings will serve as a valuable reference for both the academic and industrial communities, guiding the development of future hardware accelerators. By understanding the common strategies employed by current accelerators and identifying the unique features that contribute to their performance, we can inform the design of next-generation AI/ML hardware that is even more tailored to the requirements of emerging workloads. In doing so, we contribute to the ongoing quest for hardware architectures that can keep pace with the relentless advancement of AI/ML technologies.

\section{Emerging AI/ML Accelerators}

\subsection{Graphcore IPU}

\textbf{System Information.}
The Graphcore Intelligence Processing Unit (IPU) system used in the test is the IPU-POD16 with four IPU-M2000 units~\cite{graphcore_datasheet}. The IPU-POD16 utilizes a Dell R6525 Poplar server with dual-socket AMD Epyc2 CPUs as the host server. Each IPU-M2000 unit (1U) comprises four GC200 IPU chips connected through IPU-Link with a bandwidth of 192 GB/s. The GC200 IPU chip contains 1472 independent IPU-tiles and can process up to 8832 separate program threads in parallel.

\begin{figure*}[t]
    \centering
      \includegraphics[width=1\linewidth]{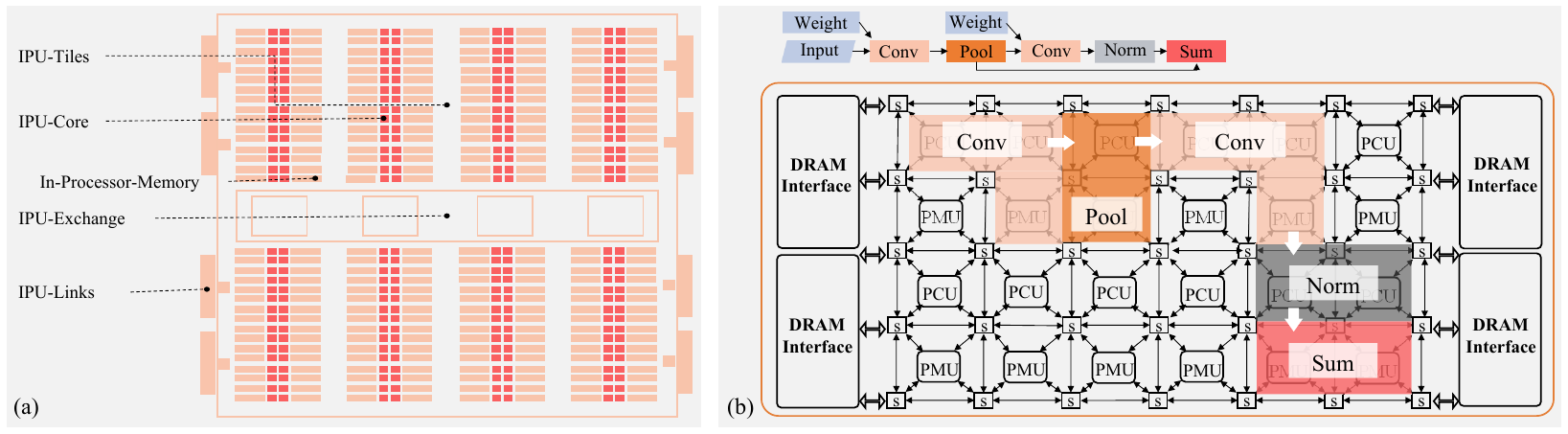}
  \caption{(a) Graphcore IPU architecture. (b) Sambanova RDU architecture}
    \label{fig:IPU_GC200}
\end{figure*}

\textbf{Memory Resource and Hierarchy.}
The architecture of the GC200 IPU chip is illustrated in Fig.~\ref{fig:IPU_GC200}(a). Each tile of the GC200 IPU chip~\cite{graphcore_GC200} is equipped with 624 KiB of local scratchpad memory, and 1472 tiles contribute to a total of 900 MB of in-processor memory with 47.5 TB/s on-chip bandwidth for a single chip. Columns of IPU-tiles are connected through IPU-Exchange with 8 TB/s bandwidth \cite{graphcore_GC200}, but with higher latency penalty \cite{jia2019dissecting}. Unlike cache, the scratchpad memory within an IPU-tile can be accessed irregularly without penalty. Each GC200 IPU chip features 10 IPU-Links and supports up to 320 GB/s chip-to-chip bandwidth. In addition to the abundant on-chip SRAM memory, each IPU-M2000 unit also offers 448 GB of streaming DDR memory with 20 GB/s bandwidth, which is used to store the dataset or output and support infrequent transactions.

\textbf{Computing Resource.}
The GC200 IPU chip employs Accumulating Matrix Product (AMP) units for its floating-point computation. Each IPU-tile has one AMP unit and can perform up to 64 multiply-accumulate (MAC) operations per cycle. The theoretical computing throughput of a single GC200 IPU chip is 250 TFLOPs (with sparsity) for FP16 format and 62.5 TFLOPs (with sparsity) for FP32 format. A single IPU-M2000 unit can achieve up to 1 PetaFLOPs peak throughput for FP16 format.

\textbf{Architecture Information}
IPUs offer a much larger core count than CPU platforms, and each IPU-tile is capable of executing completely distinct programs~\cite{jia2019dissecting}. Compared to GPU platforms, IPU-tiles are connected with high-performance on-chip networks, providing greater flexibility for Multiple Instruction Multiple Data (MIMD) programming capability and better suitability for sparse and irregular processing. The IPU programming model follows the standard Bulk Synchronous Parallel (BSP) model~\cite{valiant1990bridging}. The IPU execution flow consists of three steps: (1) local compute, (2) global synchronization, and (3) on-chip data exchange. During the compute phase, IPU-tiles operate on their local data. After the computation is completed, IPU-tiles synchronize before data exchange. The data exchange phase is supported by an on-chip interconnect with efficient point-to-point and collective communications~\cite{jia2019dissecting}. Each IPU-tile has hardware support for six threads with Simultaneous Multi-Threading (SMT)~\cite{tullsen1995simultaneous} capability to use shared hardware resources in a round-robin fashion. SMT technology can effectively hide memory access and branch latency, increasing overall throughput.

\textbf{Compiler Stack \& Parallelism Support.}
The Graphcore IPU utilizes the Poplar compiler stack~\cite{graphcore_poplar} for runtime deployment and optimization. For machine learning workloads, popular frameworks such as TensorFlow and PyTorch are used as front-ends for easier development. The Poplar Advanced Run Time (PopART) serves as the interface between the machine learning front-end and the Poplar Graph Compiler. The output of the Poplar Graph Compiler is further processed by the Graph Element Compiler to exploit efficient computation and computation patterns according to the BSP model. IPUs have abundant instruction set support for various applications, and Graph Elements can be written using C/C++ with LLVM compiler or IPU assembly. The current version of IPU supports task-level parallelism~\cite{graphcore_orogram} across multiple IPUs. However, only a limited form of task parallelism can be used within an IPU chip to overlap the IO latency. Consequently, single-chip IPU execution flow optimizes individual operator performance and has limited support for operator-wise parallelism.

\subsection{Sambanova RDU}
\textbf{System Information.}
The Sambanova Reconfigurable Dataflow Unit (RDU) system SN10-8 used in the test is built upon eight Cardinal SN10 RDUs~\cite{olukotun2020plasticine,prabhakar2022sambanova}. The SN10 RDU adopts a coarse-grain reconfigurable array (CGRA) data-flow architecture \cite{hartenstein2001coarse} for its hardware. As shown in Fig.~\ref{fig:IPU_GC200}(b), the basic units of RDU~\cite{prabhakar2018plasticine} include pattern compute units (PCUs), pattern memory units (PMUs), and pipelined switches for the on-chip network. Each PCU is equipped with six SIMD stages, and each stage has 16 single instruction multiple data (SIMD) lanes, totaling 96 Functional Units (FUs). Each SN10 chip contains 640 PCUs and 640 PMUs, providing 320 MB of on-chip SRAM. The PCU for the SN10 chip supports only the FP16 format for arithmetic computations, and the theoretical throughput of a single SN10 chip is 325 TFLOPs.

\textbf{Memory Resource and Hierarchy.}
Each PMU scratchpad of the SN10 has 512 KB of memory and 16 banks, with the number of banks matching the number of PCU lanes to provide vectorized data access. The PMU supports several memory access modes~\cite{prabhakar2018plasticine} to facilitate different applications: strided-banking mode for dense operators, FIFO mode for streaming data, line-buffer mode for sliding window accesses, duplication mode to support parallel reads, and N-buffering mode for coarse-grained pipelines. In addition to on-chip memory resources, the SN10 can be connected to up to 1.5 TB of external DRAM for streaming and random accesses. The streaming access of external DRAM supports up to 100 GB/s bandwidth, while random (sparse) access supports only 12 GB/s bandwidth.

\textbf{Architecture Information.}
The data-flow graph is mapped across PCU and PMU stages, utilizing vectorization to increase the parallelism level within each PCU lane. Each PCU has multiple SIMD lanes and stages, exploiting fine-grained parallelism for its computation tasks. The PCU consumes pipelined data from previous stages and generates pipelined data for later stages. Communication between PCUs and PMUs is based on streaming to avoid pipeline stalls and memory latency overhead. Reconfigurable controller blocks are distributed among PCUs and PMUs to match data stream rates and trigger instruction execution. Operators within the data-flow graph are mapped to multiple PCUs and PMUs depending on their size. The hardware mapping and scheduling of PCUs and PMUs to the data-flow graph aim to maximize throughput and minimize latency.

\textbf{Compiler Stack \& Parallelism Support.}
The Sambanova system utilizes SambaFlow~\cite{sambaflow} as the end-to-end compiler framework for machine learning acceleration. SambaFlow takes open-source frameworks such as PyTorch and TensorFlow as entry points to build the DNN model. The model is then fed into a dataflow graph analyzer to generate a dataflow graph with Spatial intermediate representation (IR)~\cite{koeplinger2018spatial}. The dataflow graph analyzer conducts design space exploration and performs domain-specific optimizations, such as layer fusion for DNN model mapping on RDU hardware. If the operators are not present in existing ML frameworks, users can also specify new operators through the tensor index notation API. The template compiler maps the operator into an optimized dataflow implementation, called a spatial template, on RDU hardware. The final dataflow compiler and assembler layer performs final transformations such as parallelization, placement, and routing. The dataflow graph is then transformed into the final RDU hardware mapping, and the executable file is generated.





\begin{figure*}[h!]
    \centering
\begin{multicols}{5}
\subfloat [\label{fig:Graphcore_squareMM}Graphcore]   {\includegraphics[height=0.79\columnwidth]{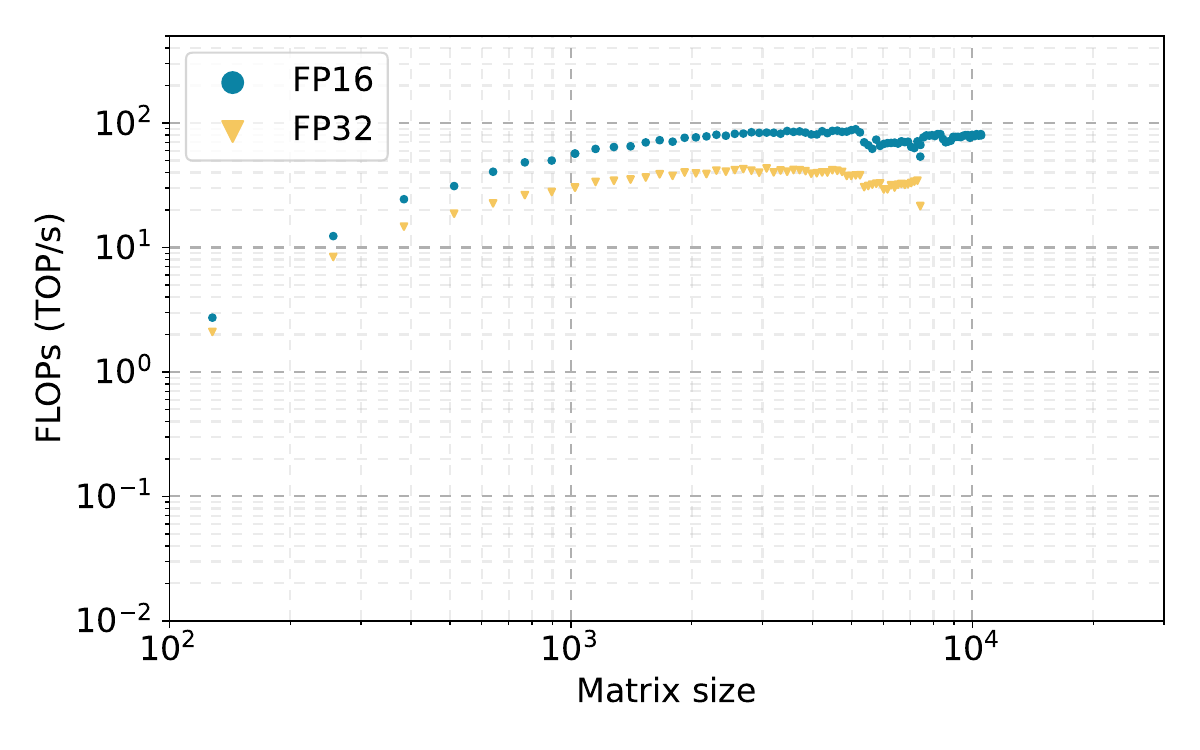}\par }
\subfloat  [\label{fig:Sambanova_squareMM}Sambanova]  {\includegraphics[height=0.79\columnwidth]{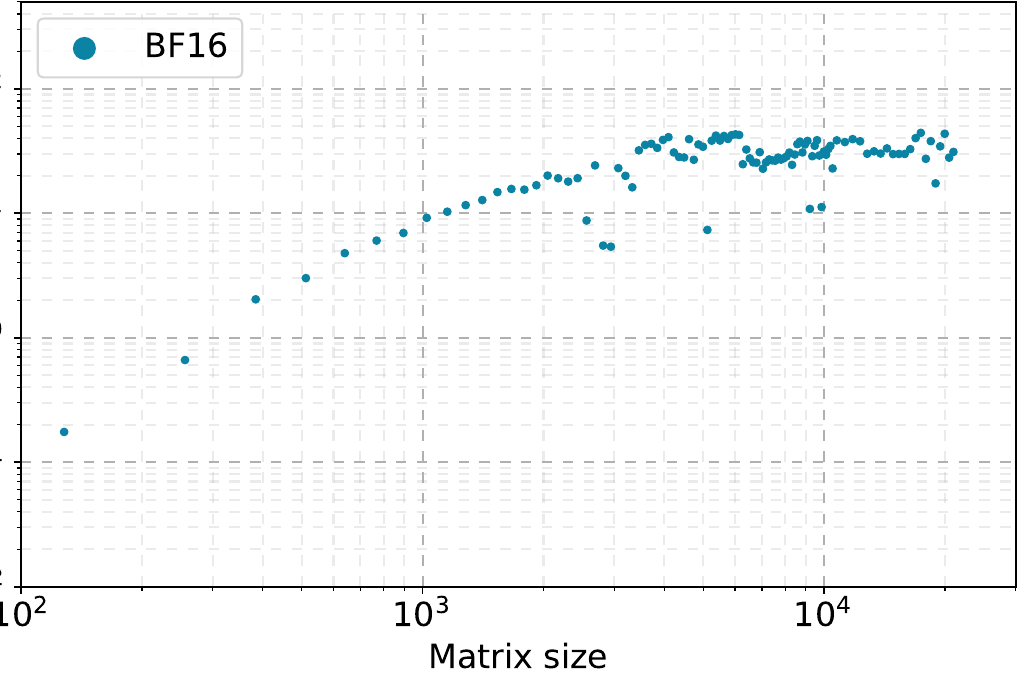}\par }
\subfloat  [\label{fig:MI100_squareMM}MI100]  {\includegraphics[height=0.79\columnwidth]{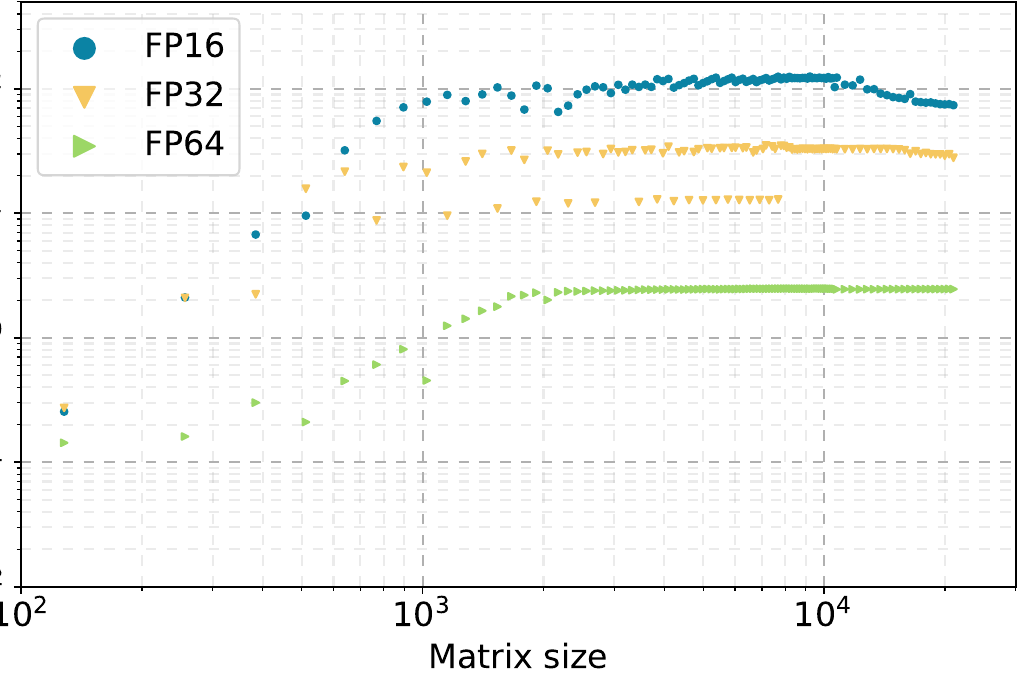}\par }
\subfloat  [\label{fig:V100_squareMM}V100]  {\includegraphics[height=0.79\columnwidth]{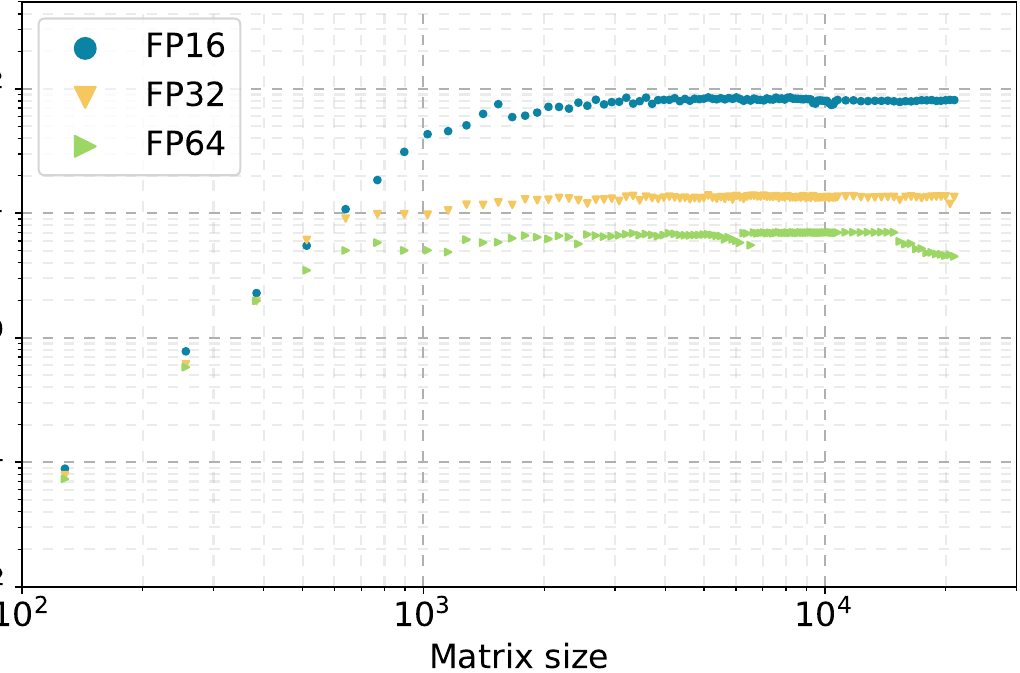}\par }
\subfloat  [\label{fig:A100_squareMM}A100]  {\includegraphics[height=0.79\columnwidth]{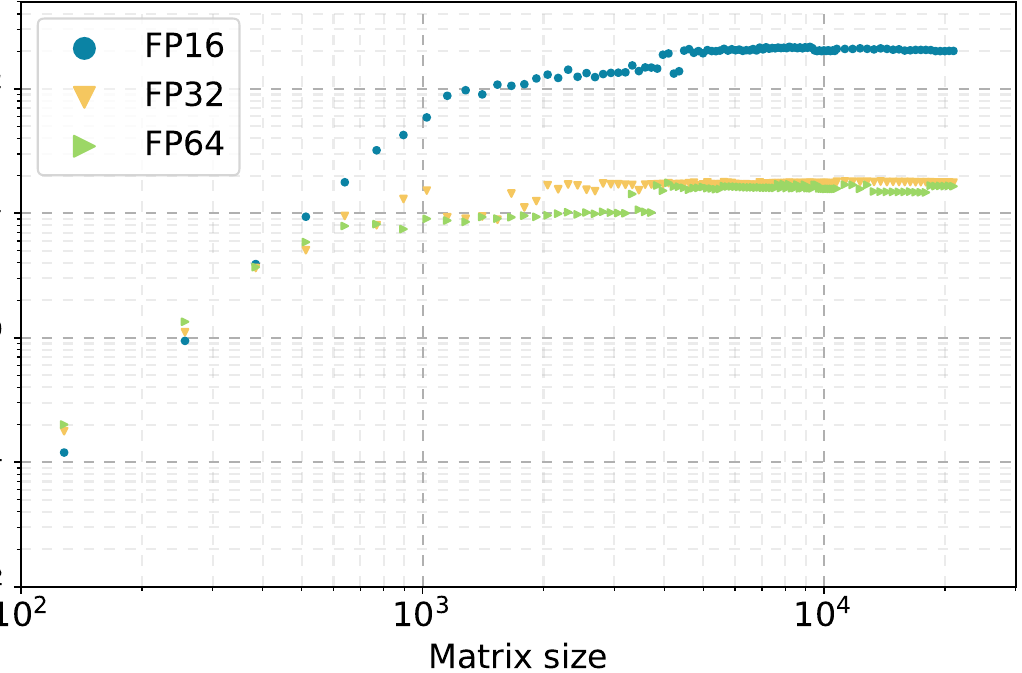}\par }
\end{multicols}
\centering
    \caption{Cross platform evaluation on square GEMM operators.}
    \label{fig:squareMM}
\end{figure*}

\subsection{Nvidia \& AMD GPU}

\textbf{System Information.}
We use Nvidia V100~\cite{Nvidia_V100} and A100~\cite{Nvidia_A100}, as well as AMD MI100 and MI250 for the GPU platform benchmark evaluation. Detailed information such as L1 cache, L2 cache, global memory bandwidth, and theoretical throughput is provided in Tab.~\ref{tab:hardware_info}.

\textbf{Nvidia GPU Architecture}
For Nvidia GPUs, the streaming multiprocessors (SM) serve as the basic unit for instruction execution and scheduling. An SM may have multiple SM partitions, which share the same global L1 data cache (shared memory) and L1 instruction cache. Each SM partition has its own register file, L0 instruction cache, warp scheduler, dispatch unit, and computing resources.
Computing resources within the SM partition include floating-point Compute Unified Device Architecture (CUDA) cores and Tensor Cores. Tensor Cores perform multiple FP16/FP32 mixed-precision fused multiply-add (FMA) operations within a single cycle and offer much higher computational throughput. Thirty-two threads are grouped into a warp and are mapped to a single SM partition for SIMD execution. Warp execution is overlapped through a fine-grained pipeline to hide instruction fetch or memory fetch latency. Multiple SM units are grouped into a single texture processing cluster (TPC), and multiple TPCs are grouped into a single GPU processing cluster (GPC). GPCs share a global L2 cache for on-chip data buffering.

\textbf{AMD GPU Architecture}
AMD GPUs use the CDNA architecture~\cite{AMD_CDNA,AMD_CDNA2} for their MI100 and MI250 products. The architecture is similar to the Nvidia Volta/Ampere architecture. The CDNA architecture employs terms such as Compute Unit (CU), CU Core, and Matrix Core, as opposed to SMs, CUDA core, and tensor core for Nvidia GPUs. The CU uses the Graphics Core Next (GCN) architecture \cite{AMD_GCN} and has multiple floating-point cores and a single Matrix Core with enhanced throughput.

\textbf{Memory Resource and Hierarchy.}
To boost the data transfer rate between GPU on-chip resources and off-chip memory, High Bandwidth Memory (HBM) technology is widely adopted for both Nvidia and AMD GPUs. HBM consists of stacks of DRAM dies with through-silicon via (TSV) connections, providing up to several TB/s of external memory bandwidth. The connection of HBM dies with GPU dies is based on an underlying silicon interposer to ensure reliable and high-speed connections.

\textbf{Programming Model.}
Nvidia employs the CUDA parallel programming model~\cite{Nvidia_CUDA} for its GPUs to leverage their massive processing power with minimal implementation coding effort. It enables heterogeneous computation where the CPU and GPU have separate memory and thread spaces. The CUDA programming is based on kernel functions, which are called by the host CPU and executed on the GPU device. The kernel function is executed with massive concurrent threads on the GPU that share the same kernel code. The thread ID is used for memory addressing and thread cooperation. CUDA threads are extremely lightweight, allowing them to be created or switched with minimal penalties. Thirty-two threads are grouped into a single warp, and the warp scheduler determines the sequence of warp execution to hide instruction and memory access latency. To reduce global memory traffic, thread cooperation within a thread block is enabled through shared memory (L1 cache) and thread synchronization primitives. An L2 cache with residency control is also available to further reduce global memory access bandwidth and latency. Thread blocks to SMs mapping are scheduled during runtime, and multiple thread blocks can be mapped to a single SM. The CUDA programming model makes the CUDA code scalable for various hardware platforms, ranging from laptops to high-end servers with minimal changes.

AMD adopted ROCm~\cite{AMD_rocm} for its software stack and uses the Heterogeneous Interface for Portability (HIP)~\cite{AMD_hip} programming model. HIP Runtime API and kernel programming methods can be used for both AMD and Nvidia GPU platforms. The HIP programming model is similar to the CUDA programming model.

\textbf{Compiler Support}
Existing deep learning frameworks, such as TensorFlow and PyTorch, can be used with the CUDA and ROCm software stack to support automatic differentiation and DNN computational graph generation.

\begin{table}[h!]
\caption{System information.}
\begin{adjustbox}{width=\columnwidth}

\begin{tabular}{cccc}
\hline
Device           & SambaNov SN10                                                             & GraphCore GC200                                                          & NvidiaV100                                                          \\ \hline \hline
Host CPU         & AMD EPYC 7742                                                             & AMD EPYC 7302                                                            & Intel E5-2620                                                       \\ \hline
Software   Stack & \begin{tabular}[c]{@{}c@{}}PyTorch 1.10.2\\ SambaFlow 1.14.0\end{tabular} & \begin{tabular}[c]{@{}c@{}}TensorFlow 1.15.5\\ Poplar 2.4.0\end{tabular} & \begin{tabular}[c]{@{}c@{}}PyTorch 1.12.1\\ CUDA 11.6\end{tabular}  \\ \hline \hline
Device           & Nvidia A100                                                               & AMD MI100                                                                & AMD MI250                                                           \\ \hline \hline
Host CPU         & AMD EPYC 7502                                                             & AMD EPYC 7543                                                            &                                                                     \\ \hline
Software   Stack & \begin{tabular}[c]{@{}c@{}}PyTorch 1.12.1\\ CUDA 11.6\end{tabular}        & \begin{tabular}[c]{@{}c@{}}PyTorch 1.13.0\\ ROCm 5.3.0\end{tabular}      & \begin{tabular}[c]{@{}c@{}}PyTorch 1.13.0\\ ROCm 5.3.0\end{tabular} \\ \hline
\end{tabular}
\label{tab:system_info}
\end{adjustbox}
\end{table}





\begin{figure*}[h!]
    \centering
\begin{multicols}{5}
\subfloat [\label{fig:Graphcore_BERTMM}Graphcore]   {\includegraphics[height=0.79\columnwidth]{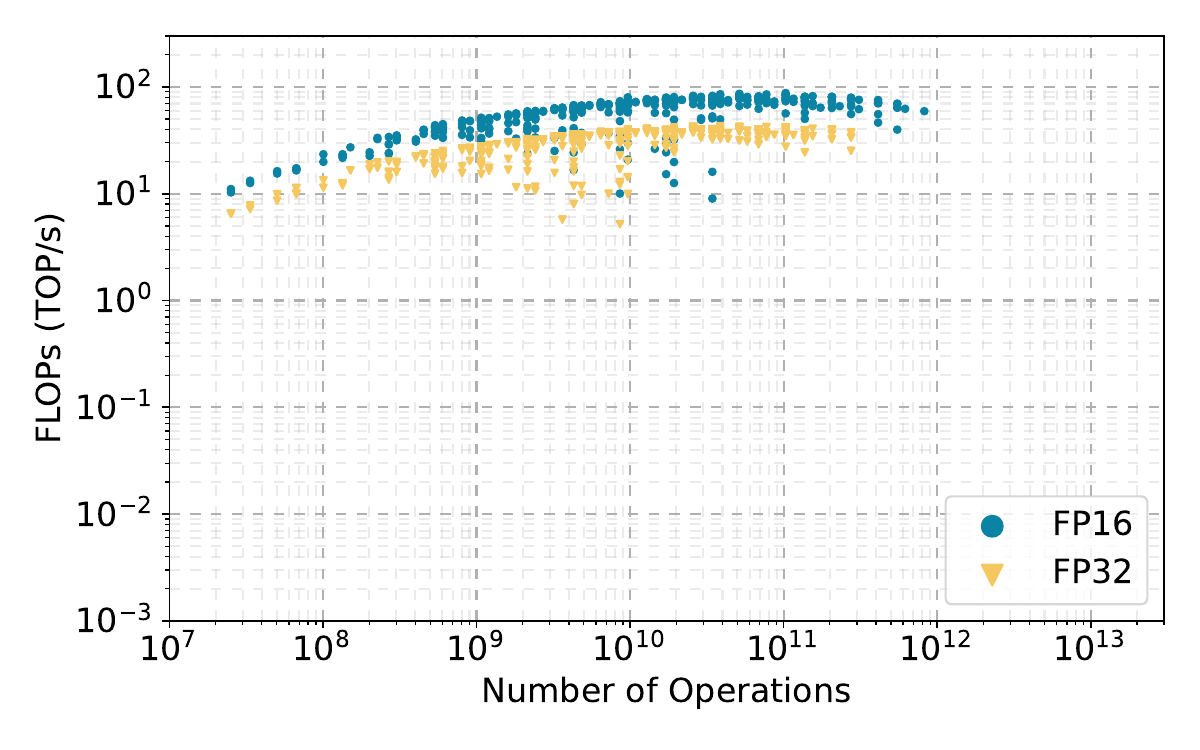}\par }
\subfloat  [\label{fig:Sambanova_BERTMM}Sambanova]  {\includegraphics[height=0.79\columnwidth]{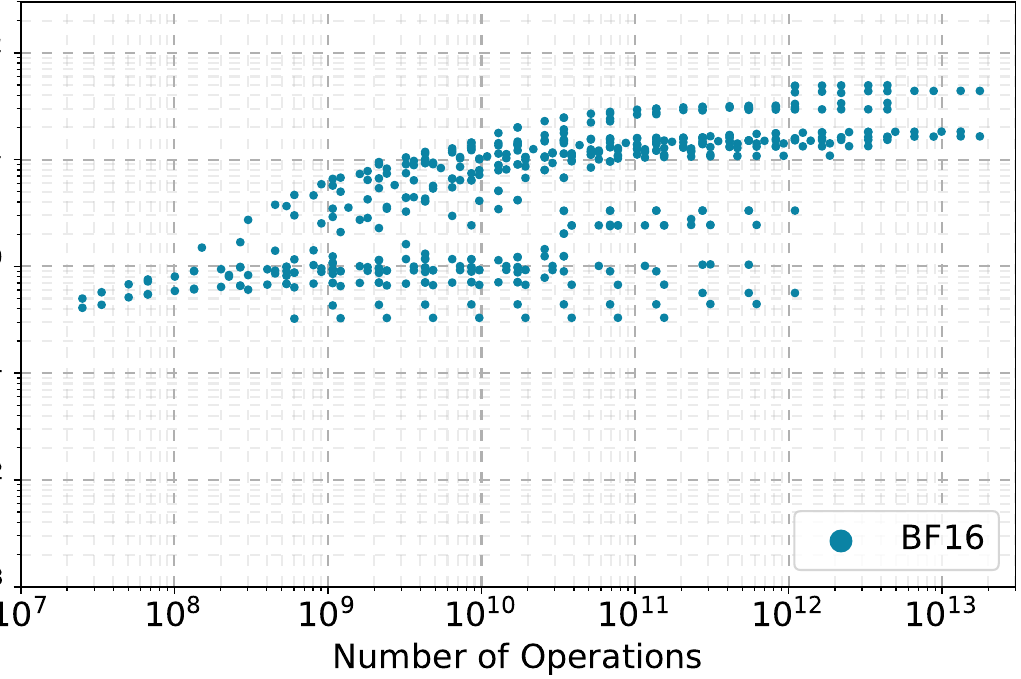}\par }
\subfloat  [\label{fig:MI100_BERTMM}MI100]  {\includegraphics[height=0.79\columnwidth]{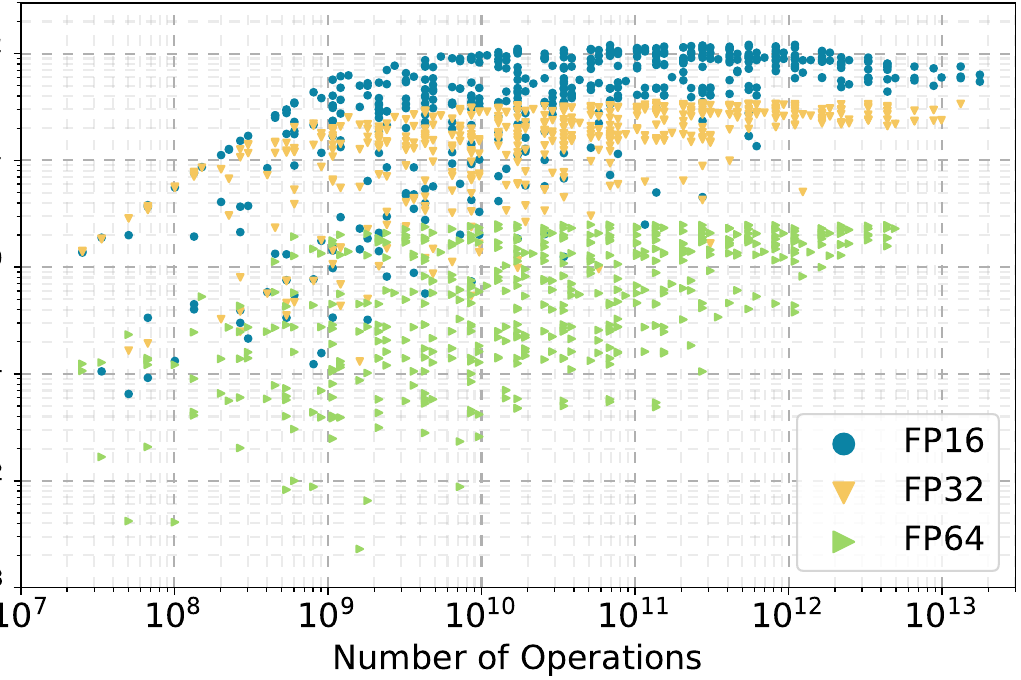}\par }
\subfloat  [\label{fig:V100_BERTMM}V100]  {\includegraphics[height=0.79\columnwidth]{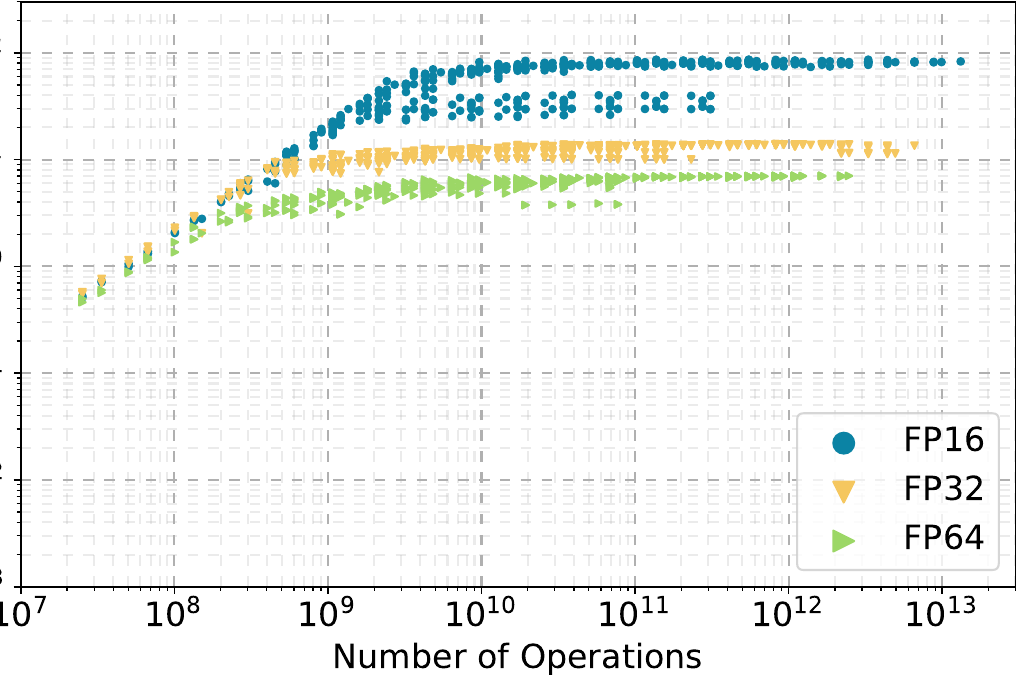}\par }
\subfloat  [\label{fig:A100_BERTMM}A100]  {\includegraphics[height=0.79\columnwidth]{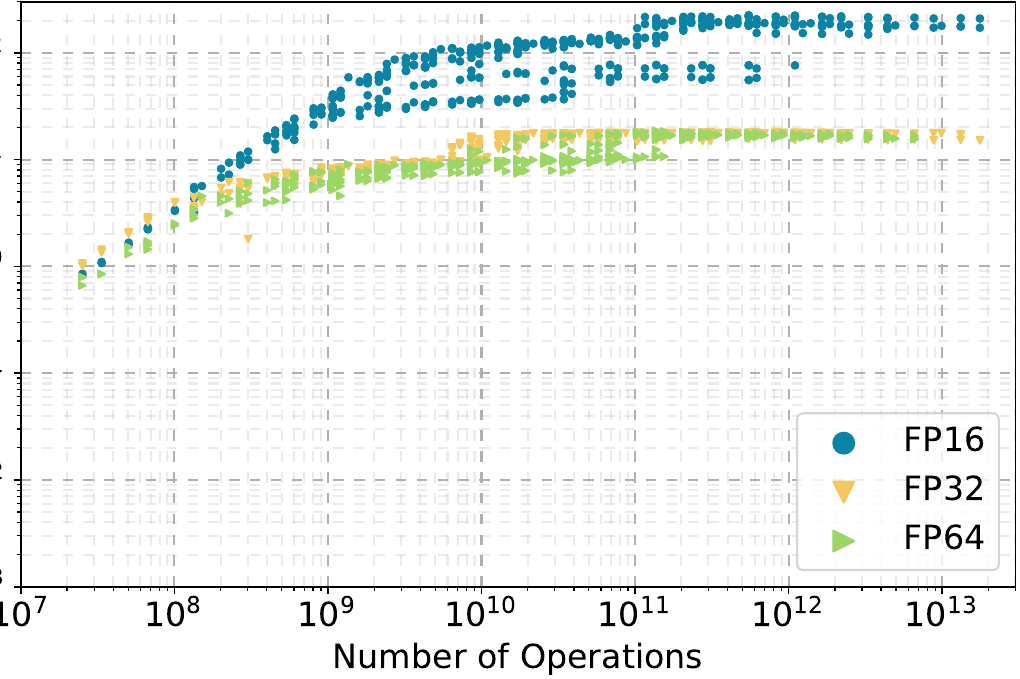}\par }
\end{multicols}
\centering
    \caption{Cross platform evaluation on BERT operators.}
    \label{fig:BERTM}
\end{figure*}





\begin{figure*}[h!]
    \centering
\begin{multicols}{5}
\subfloat [\label{fig:Graphcore_Conv}Graphcore]   {\includegraphics[height=0.79\columnwidth]{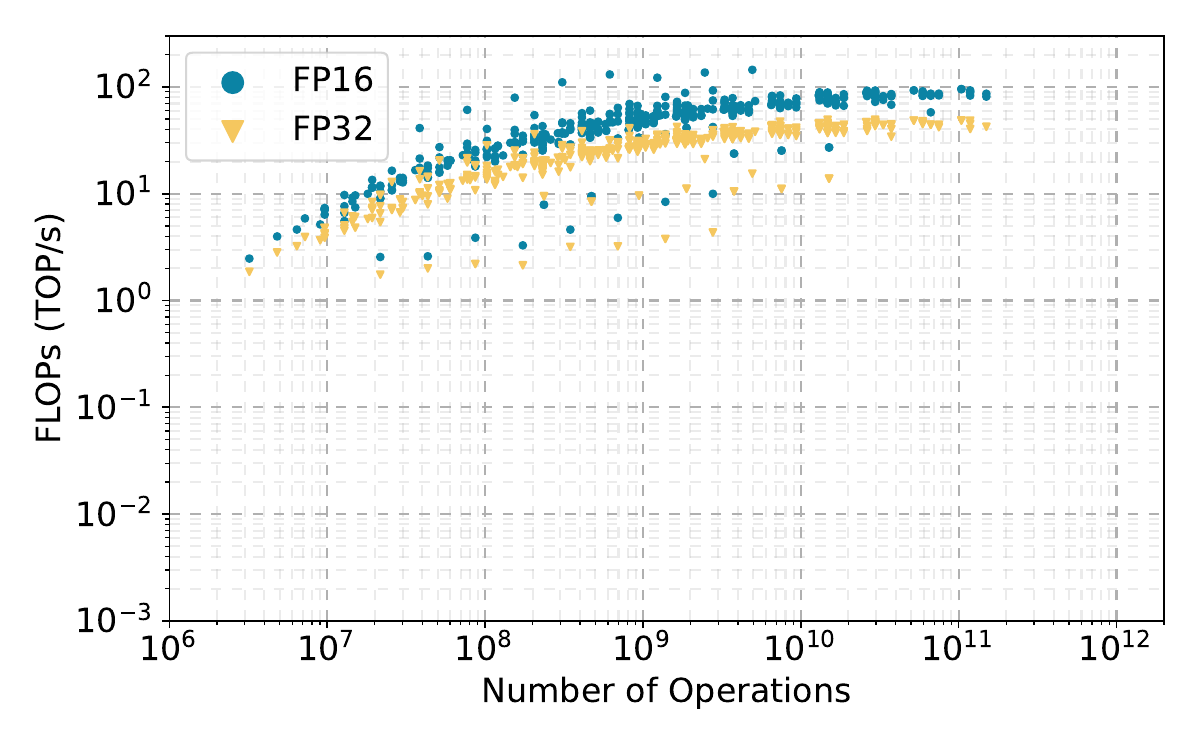}\par }
\subfloat  [\label{fig:Sambanova_Conv}Sambanova]  {\includegraphics[height=0.79\columnwidth]{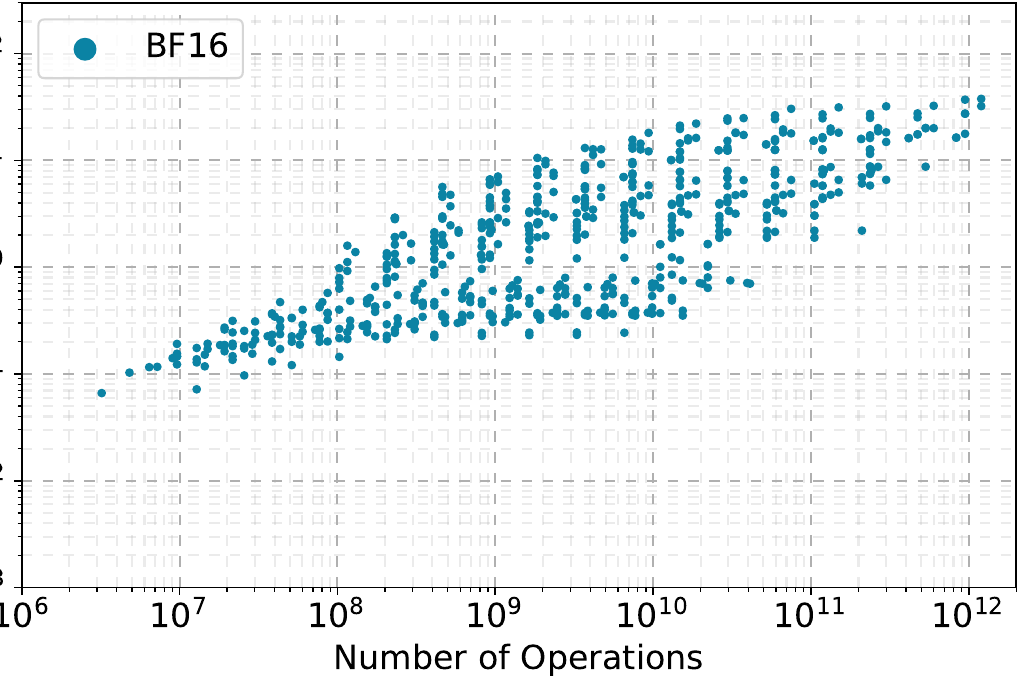}\par }
\subfloat  [\label{fig:MI100_Conv}MI100]  {\includegraphics[height=0.79\columnwidth]{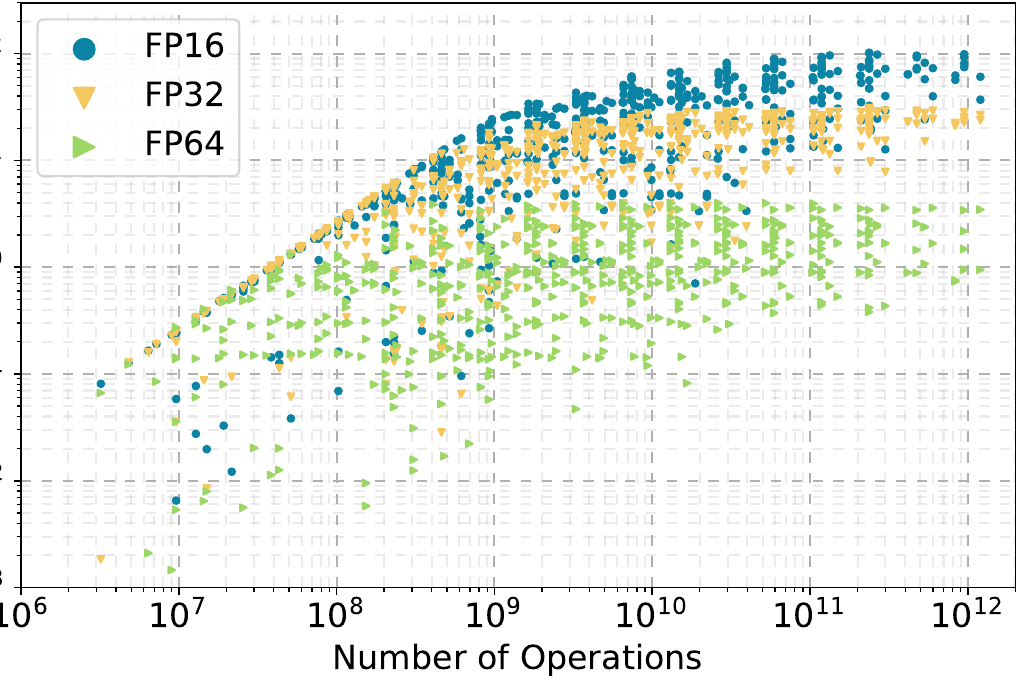}\par }
\subfloat  [\label{fig:V100_Conv}V100]  {\includegraphics[height=0.79\columnwidth]{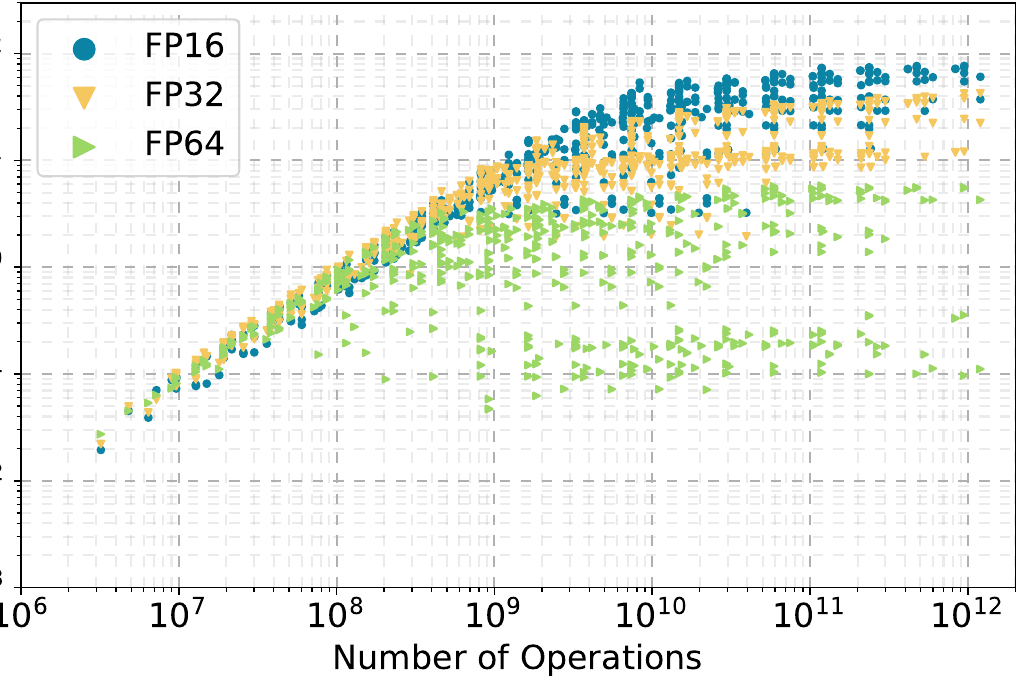}\par }
\subfloat  [\label{fig:A100_Conv}A100]  {\includegraphics[height=0.79\columnwidth]{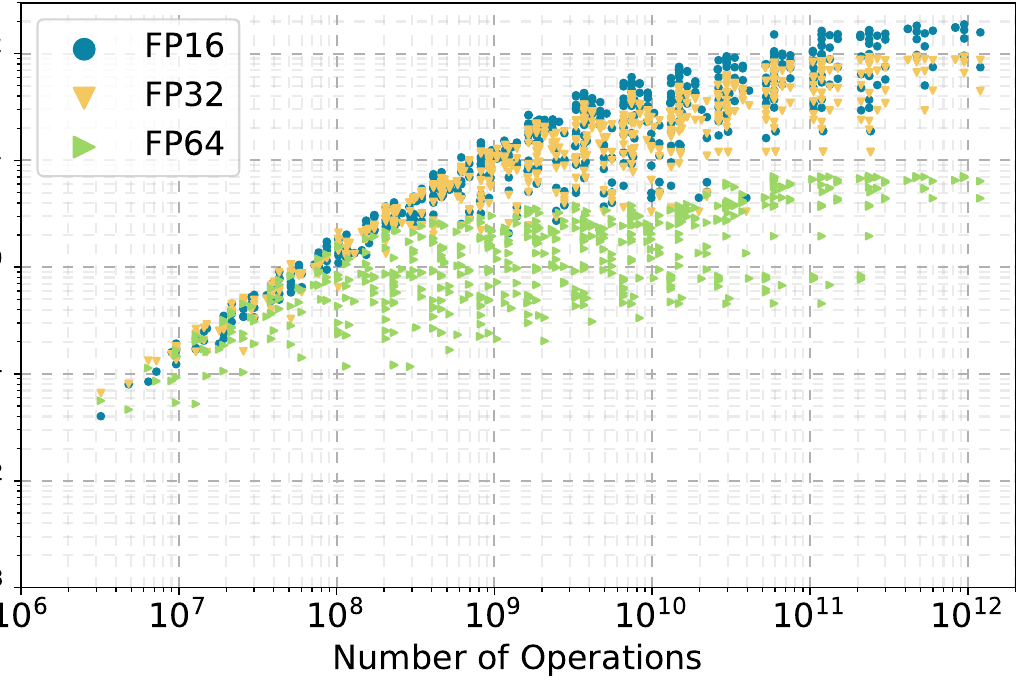}\par }
\end{multicols}
\centering
    \caption{Cross platform evaluation on 2D convolution operators.}
    \label{fig:Conv}
\end{figure*}

\subsection{Summary}

\textbf{Similarity Between Platforms}
Graphcore IPU, Sambanova RDU, and GPU platforms each have their own basic SIMD/SIMT units: IPU-tile, PCU, and SM. These basic SIMD units share similar architecture and process lightweight concurrent threads while switching between thread groups (warps) with minimal penalty. Such SIMD architecture fully exploits fine-grained pipelined parallelism, enabling high throughput and high energy-efficient parallel computation.





\begin{figure*}[h!]
    \centering
\begin{multicols}{4}
\subfloat [\label{fig:Graphcore_elem}Graphcore]   {\includegraphics[height=0.81\columnwidth]{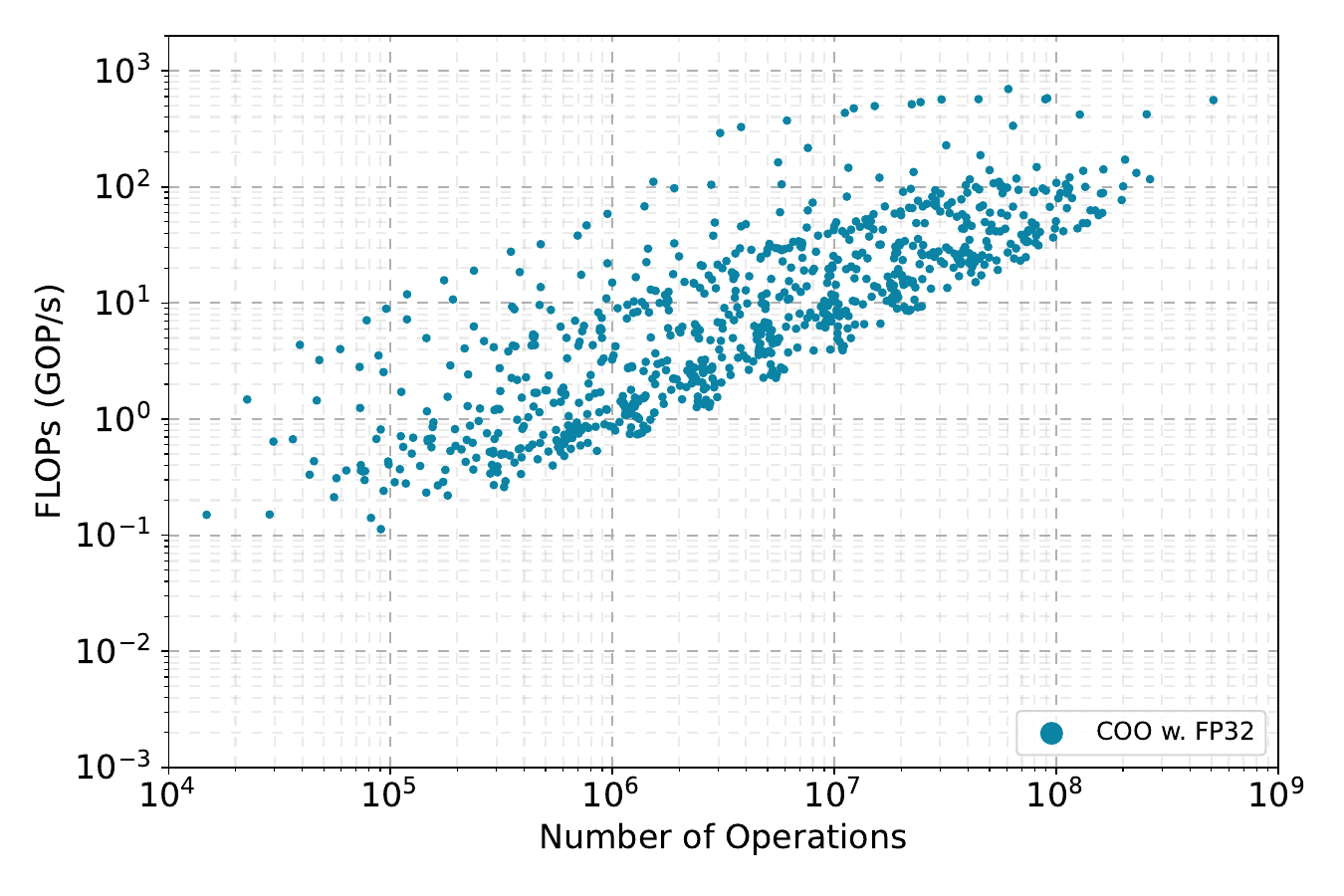}\par }
\subfloat  [\label{fig:MI100_elem}MI100]  {\includegraphics[height=0.81\columnwidth]{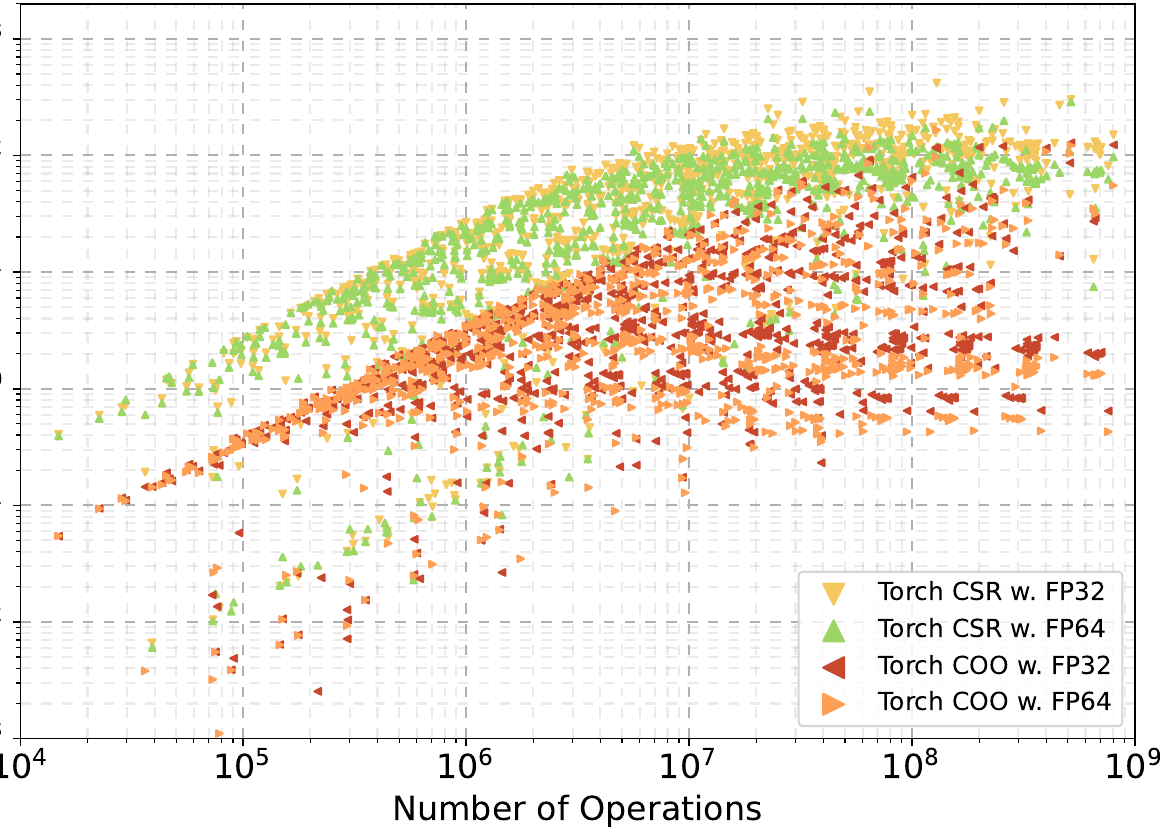}\par }
\subfloat  [\label{fig:V100_elem}V100]  {\includegraphics[height=0.81\columnwidth]{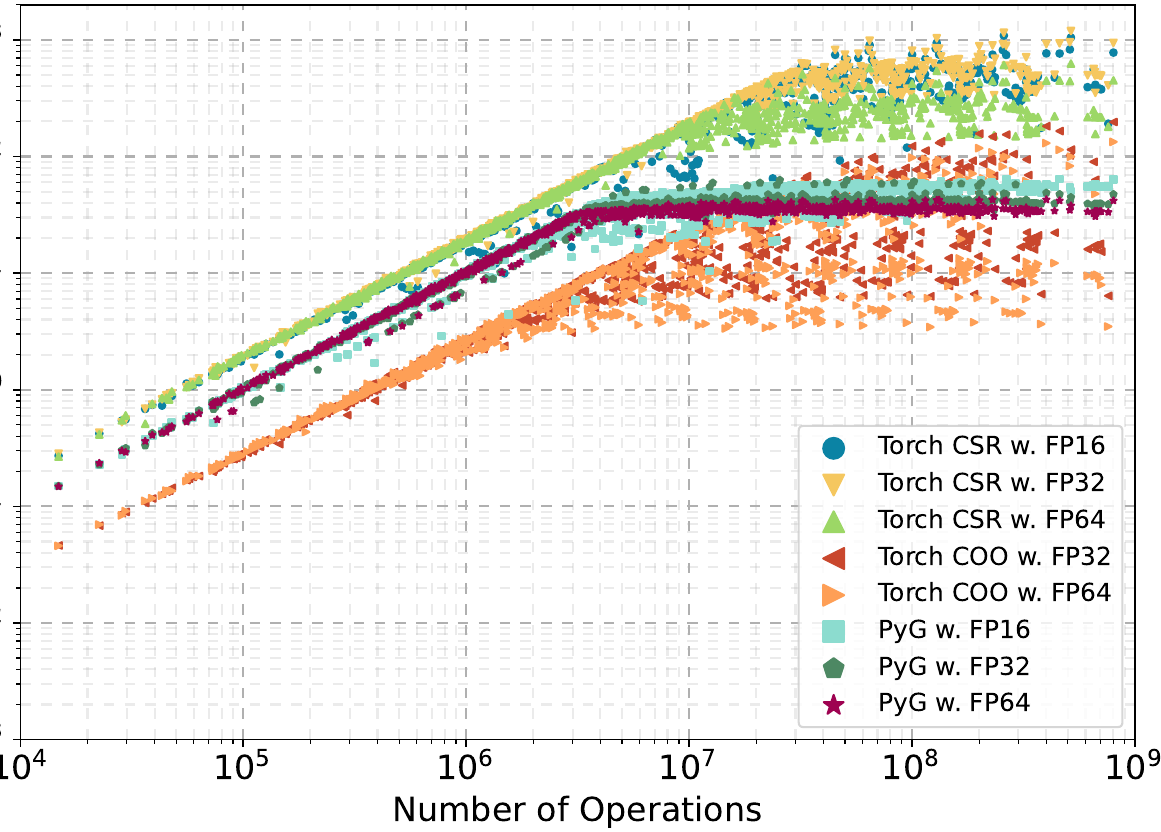}\par }
\subfloat  [\label{fig:A100_elem}A100]  {\includegraphics[height=0.81\columnwidth]{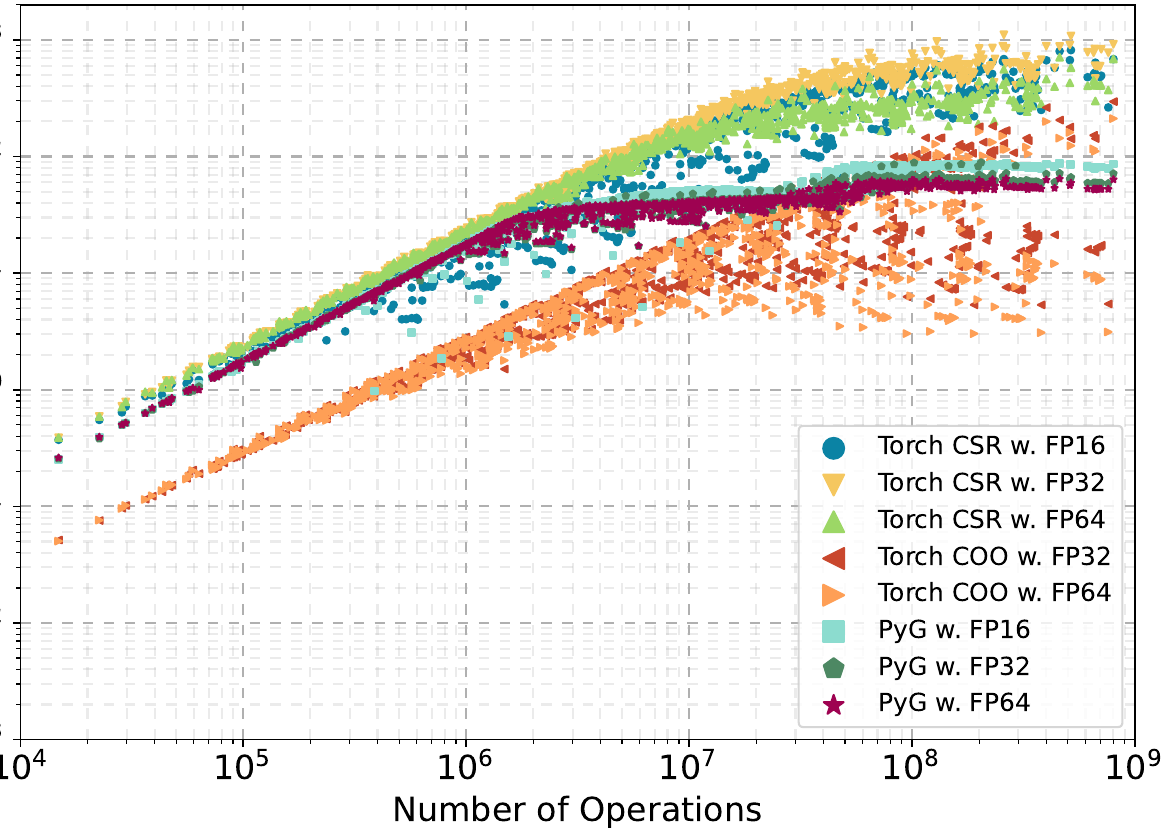}\par }
\end{multicols}
\centering
    \caption{Cross platform evaluation on SPMM operators.}
    \label{fig:spmm}
\end{figure*}

\textbf{Difference Between Platforms}
Although the three accelerator platforms' basic SIMD/SIMT units have similar architecture and execution models, the interconnect and scheduling between these units differ among platforms. Graphcore IPU features a global crossbar that connects every IPU-tile, and the scheduling and execution model between IPU-tiles follows the BSP model with three steps: local computation, synchronization, and data communication/exchange. Sambanova RDU provides reconfigurable switches for PCU and PMU connections, where scheduling and mapping of PCUs and PMUs use the Spatial dataflow graph compiler for task parallelism. GPU architecture has limited connections and communication bridges between SMs, and the execution and mapping of tasks to the GPU follow the CUDA programming model with SIMT style. Overall, Graphcore IPU and Sambanova IPU platforms offer greater flexibility in SIMD/SIMT unit mapping and scheduling than the GPU platform, potentially providing more advantages for applications with sparse or irregular computation and communication patterns.

\section{Evaluation Results}
\label{sec:setting}

We conduct benchmark evaluations of several commonly used DNN operators on the target platforms, including \ding{182} general matrix multiplication (GEMM), \ding{183} 2D convolution (Conv2D), and \ding{184} sparse-matrix dense-matrix multiplication (SPMM). It is important to note that the Sambanova SN10 platform lacks compiler support for SPMM, and therefore is not evaluated for this operator.



\subsection{Square GEMM Benchmark}
We conduct benchmarking of square matrix multiplication following the software setup outlined in Table~\ref{tab:system_info}. The matrix size is varied from 256 to 10,624 with a 128 step size, and from 10,752 to 20,992 with a 512 step size. We record throughput performance over available floating-point formats for each platform. The Graphcore GC200 IPU platform is benchmarked for its available FP16 and FP32 data types, with matrix size limited to 10,496 due to limited on-chip memory. Both FP16, FP32, and FP64 throughputs were recorded for the GPU platforms. The evaluation results are shown in Fig.~\ref{fig:squareMM}. It is worth noting that for the Sambanova platform, throughput performance becomes unstable when matrix sizes grow to 2500 and above, due to unstable PCUs and PMUs mapping through the SambaFlow compiler. The AMD MI100 GPU platform shows unstable performance between matrix sizes 256 to 7936 for FP32 format due to choices of CU core and Matrix core during PyTorch to ROCm compilation. 

\begin{figure*}[h!]
    \centering
\begin{multicols}{5}
\subfloat [\label{fig:Graphcore_elem}Graphcore]   {\includegraphics[height=0.79\columnwidth]{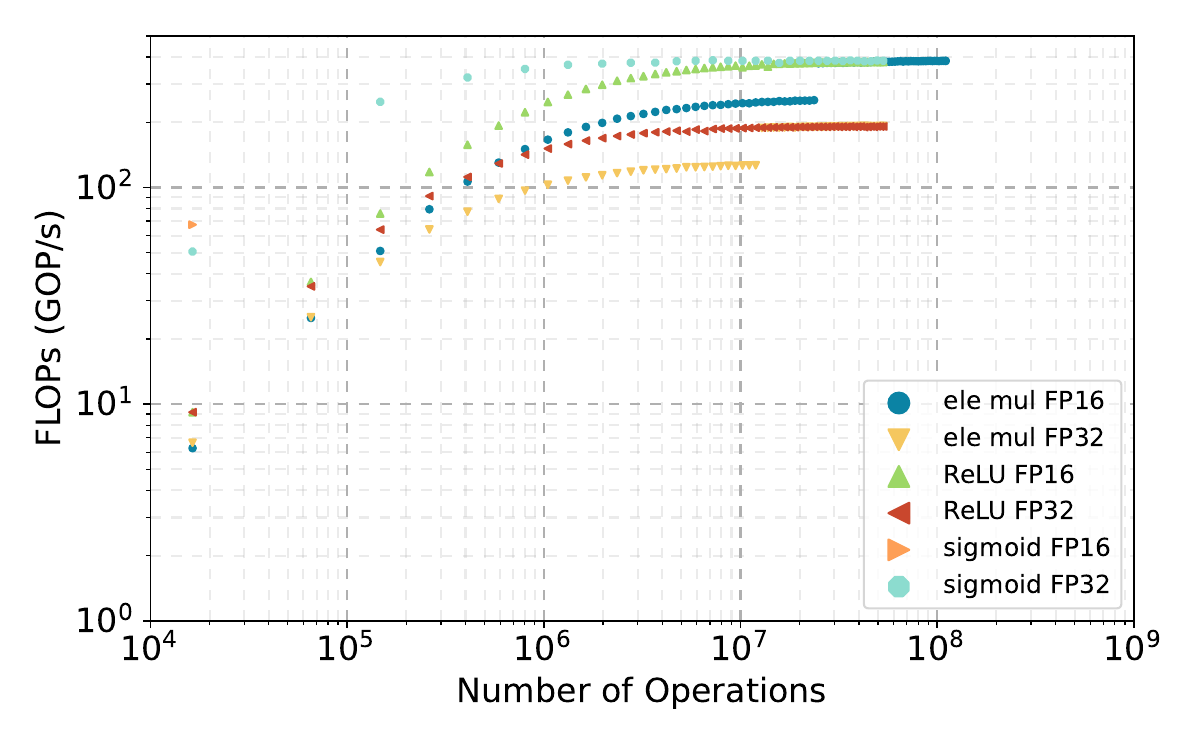}\par }
\subfloat  [\label{fig:Sambanova_elem}Sambanova]  {\includegraphics[height=0.79\columnwidth]{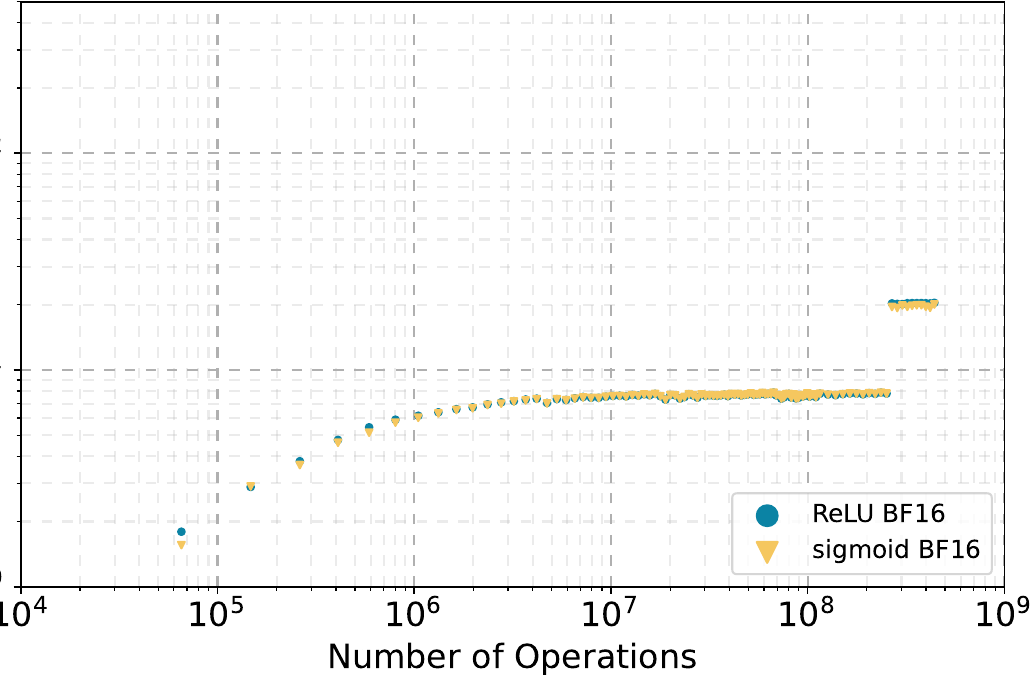}\par }
\subfloat  [\label{fig:MI100_elem}MI100]  {\includegraphics[height=0.79\columnwidth]{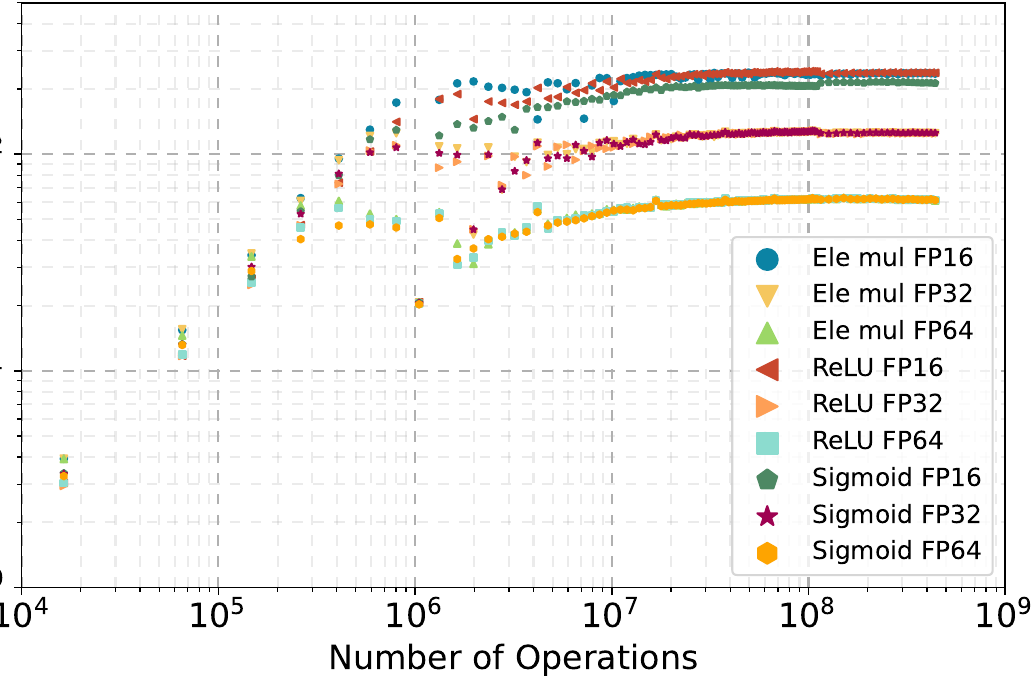}\par }
\subfloat  [\label{fig:V100_elem}V100]  {\includegraphics[height=0.79\columnwidth]{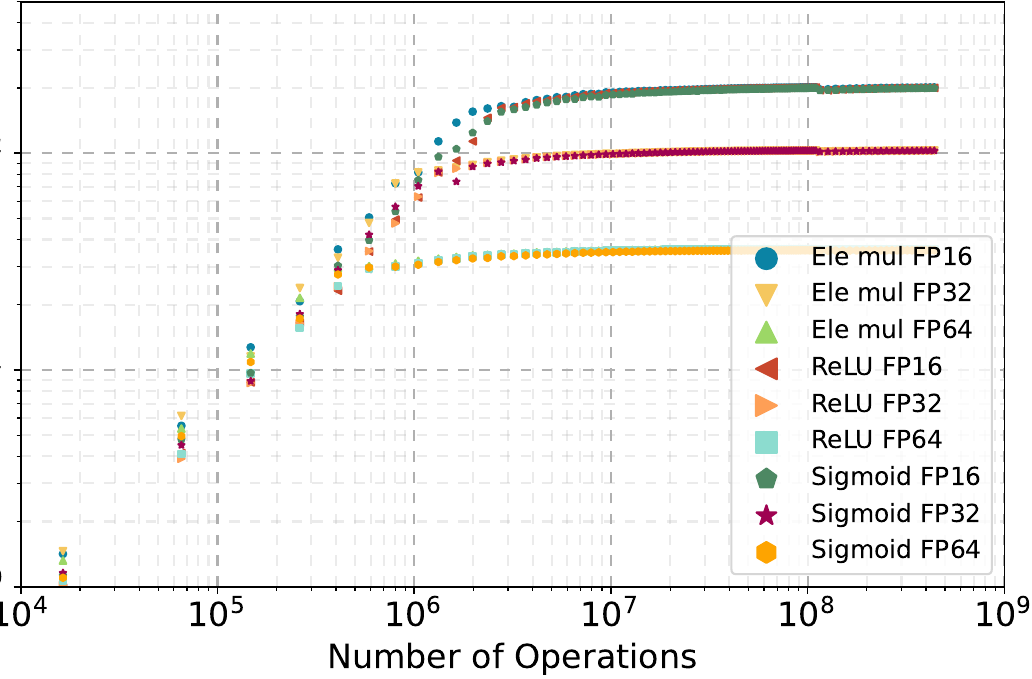}\par }
\subfloat  [\label{fig:A100_elem}A100]  {\includegraphics[height=0.79\columnwidth]{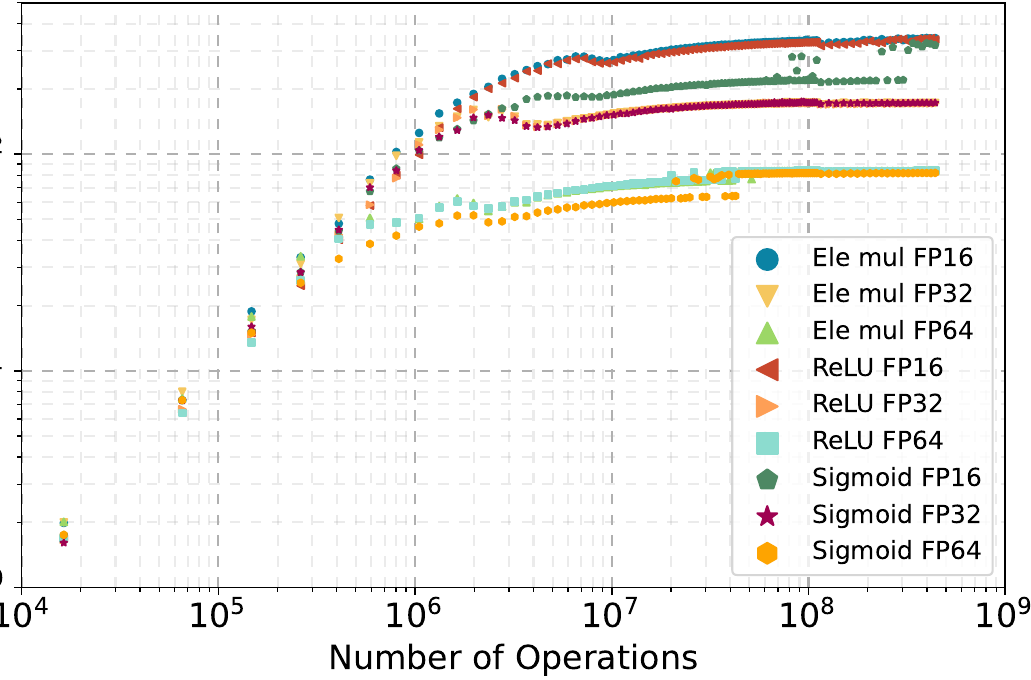}\par }
\end{multicols}
\centering
    \caption{Cross platform evaluation on element-wise operators.}
    \label{fig:elem}
\end{figure*}

\begin{table}[h!]
\caption{Normalized geomean hardware throughput for GEMM.}
\centering
\begin{adjustbox}{width=0.75\columnwidth}
\centering
\begin{tabular}{c|ccccc}
\hline
Platform & A100  & V100 & MI100 & GC200            & SN10             \\ \hline
FP16     & 10.64 & 4.81 & 7.00  & 6.00             & 2.96             \\ \hline
FP32     & 1.30  & 1.00 & 2.06  & 2.55             & \textbackslash{} \\ \hline
FP64     & 1.12  & 0.52 & 0.20  & \textbackslash{} & \textbackslash{} \\ \hline
\end{tabular}
    
\label{tab:GEMM_result}
\end{adjustbox}
\end{table}

We present the performance summary of platforms in Table~\ref{tab:GEMM_result}, which is computed by taking the geometric mean of benchmarks and using V100 FP32 format as the normalization baseline. Among all platforms, Graphcore IPU GC200 delivers the highest FP32 performance, while A100 achieves the highest FP16 and FP64 performance.

\subsection{BERT Operator Benchmark}
We benchmark the BERT model's operators across a variety of batch sizes, sequence lengths, and the number of heads. For the Graphcore GC200 IPU platform, there are 8 configurations for the batch size: [1, 2, 4, ..., 128], and 8 configurations for (sequence length, number of head) pairs: [(128, 12), (128, 16), (128, 32), (384, 12), (384, 16), (384, 32), (512, 16), (512, 32)]. For other platforms, there are 11 configurations for the batch size: [1, 2, 4, ..., 1024], while the (sequence length, number of head) configurations remain the same as those in the IPU platform setup. Benchmark results are presented in Fig.~\ref{fig:BERTM}.





\subsection{2D Convolution Operator Benchmark}
We benchmark operator performance on three commonly used convolutional neural networks (CNNs) with the same input image size as the ImageNet dataset \cite{krizhevsky2017imagenet}: ResNet-18, ResNet-50 \cite{he2016deep}, and MobileNetV2 \cite{sandler2018mobilenetv2}. We evaluate the CNN operators on eight batch size configurations (1, 2, 4, ..., 128) using the Graphcore GC200 IPU platform and 11 batch size configurations (1, 2, 4, ..., 1024) on other platforms. The benchmark results are presented in Fig.~\ref{fig:Conv}.

\begin{table}[h!]
\caption{Normalized geomean hardware throughput for convolution.}
\centering
\begin{adjustbox}{width=0.7\columnwidth}
\centering
\begin{tabular}{c|ccccc}
\hline
Platform & A100 & V100 & MI100 & GC200            & SN10             \\ \hline
FP16     & 2.26 & 1.48 & 1.99  & 8.18             & 0.35             \\ \hline
FP32     & 1.86 & 1.00 & 1.26  & 4.60             & \textbackslash{} \\ \hline
FP64     & 0.26 & 0.16 & 0.14  & \textbackslash{} & \textbackslash{} \\ \hline
\end{tabular}
    
\label{tab:CONV_result}
\end{adjustbox}
\end{table}

The geometric mean benchmark results for CNNs are presented in Table~\ref{tab:CONV_result}, where the V100 FP32 format is used as the baseline for normalization. In both FP16 and FP32 formats, the Graphcore IPU GC200 platform outperforms all other platforms with at least a 2$\times$ speedup.

\subsection{SPMM Benchmark}

Graph neural network (GNN) is significant workload in the field of AI/ML, and their performance bottleneck is the sparse-matrix dense-matrix multiplication (SPMM)~\cite{xie2023accel}. We evaluate the performance of SPMM on both Graphcore IPU GC200 and GPU platforms. For the IPU platform, we use PopSparse library~\cite{popsparse} and follow the benchmarking example in~\cite{popsparse_github}. PopSparse currently only supports COO format for sparse matrices and provides backpropagation on both sparse and dense inputs. For GPU platform, we use PyTorch sparse library~\cite{torch_sparse} to perform SPMM, which supports two sparse matrix formats: coordinate (COO) and compressed sparse row (CSR). COO format-based PyTorch SPMM supports sparse gradient backpropagation, while CSR format does not. We also provide SPMM benchmark using PyTorch Geometric (PyG)~\cite{Fey_Lenssen_2019}, an open-source message-passing-based framework, on Nvidia GPU. 

We conduct benchmarks on a selection of 100 sparse matrices sourced from the SuiteSparse library~\cite{davis2011university}. To generate additional results, we vary the second dimension (batch size) of the dense matrix. The benchmark is performed on both the GC200 IPU platform and GPU platforms, with batch sizes ranging from 8 to 4096 and 8 to 8192, respectively. Figure~\ref{fig:spmm} shows the raw results.

\begin{table}[h!]
\caption{Normalized geomean hardware throughput for SPMM.}
\begin{adjustbox}{width=\columnwidth}

\begin{tabular}{c|ccc|ccc|cc}
\hline
Framework & \multicolumn{3}{c|}{PyG}                               & \multicolumn{3}{c|}{Torch CSR}                         & \multicolumn{2}{c}{Torch COO} \\ \hline
Platform  & FP16             & FP32             & FP64             & FP16             & FP32             & FP64             & FP32    & FP64                \\ \hline
A100      & 3.86             & 3.47             & 3.36             & 8.16             & 12.24            & 9.29             & 0.95    & 0.78                \\ \hline
V100      & 2.73             & 2.64             & 2.45             & 10.01            & 11.25            & 8.54             & 1.00    & 0.81                \\ \hline
MI100     & \textbackslash{} & \textbackslash{} & \textbackslash{} & \textbackslash{} & 3.14             & 2.64             & 0.35    & 0.29                \\ \hline
GC200     & \textbackslash{} & \textbackslash{} & \textbackslash{} & \textbackslash{} & \textbackslash{} & \textbackslash{} & 1.06    & \textbackslash{}    \\ \hline
\end{tabular}
    
\label{tab:SPMM_result}
\end{adjustbox}
\end{table}

The benchmark results' geometric mean is presented in Table~\ref{tab:SPMM_result}, with the V100 FP32 format as the normalization baseline. The PyTorch CSR on V100/A100 platform achieves the highest throughput for the SPMM benchmark. The PyG framework exhibits more stable performance for differentiable SPMM and supports more data formats. The Graphcore IPU GC200 platform shows great potential in SPMM tasks, with a high peak performance; however, the performance is not optimized for varying problem sizes.

\subsection{Streaming Operator Benchmark}
We benchmark streaming operators, including element-wise square and non-linear operators such as ReLU and Sigmoid, on both platforms. The Sambanova SN10 platform's compiler stack, SambaFlow, does not support the individual element-wise square operator; therefore, it is not recorded. The benchmark results of streaming operators are presented in Fig.~\ref{fig:elem}.

\section{Conclusion}
\label{sec:conclusion}
In this study, we evaluate and compare three major types of hardware platforms for AI/ML acceleration: Graphcore IPU, Sambanova RDU, and GPU. Through our benchmarking, we have identified the potential of the Graphcore IPU for accelerating AI/ML applications such as CNNs and GNNs. Overall, our work contributes to a better understanding of the current state-of-the-art in AI/ML hardware acceleration and provides guidance for future research in this field.

\section*{Acknowledgement}
This material is based upon work supported by the U.S. Department of Energy, Office of Science, Office of Advanced Scientific Computing Research, ”ComPort: Rigorous Testing Methods to Safeguard Software Porting”, under Award Number 78284. This work uses platforms supported by U.S. DOE Office of Science, Office of Advanced Scientific Computing Research, under award 66150: ”CENATE - Center for Advanced Architecture Evaluation”.




 
\bibliographystyle{ACM-Reference-Format}
\balance
\bibliography{ref}

\end{document}